# THE ORIGIN OF COSMIC RAYS

PETER L. BIERMANN

*Max Planck Institut für Radioastronomie*

*D-53010 Bonn, Germany*

**Abstract.** In the following we describe recent progress in our understanding of the origin of cosmic rays. We propose that cosmic rays originate mainly in three sites, a) normal supernova explosions into the interstellar medium, b) supernova explosions into stellar winds, and c) hot spots of powerful radio galaxies. The proposal depends on an assumption about the scaling of the turbulent diffusive transport in cosmic ray mediated shock regions; the proposal also uses a specific model for the interstellar transport of cosmic rays. The model has been investigated in some detail and compared to i) the radio data of OB stars, Wolf Rayet stars, radio supernovae, radio supernova remnants, Gamma-ray line and continuum emission from starforming regions, and the cosmic ray electron spectrum, ii) the Akeno air shower data over the particle energy range from 10 TeV to EeV, and iii) the Akeno and Fly's Eye air shower data from 0.1 EeV to above 100 EeV.

## 1. Introduction

Ever since their discovery in 1912 by Hess and Kohlhörster the origin of cosmic rays has been one of the main enigmas of physics; the energies of the particles observed far exceed the energies even in the most powerful planned accelerators on earth. An important recent book is by Berezinskii *et al.* (1990). Discovering a clue about the origin of cosmic rays may help us to explore very high energy physics of particles as well as cosmic accelerators.

Most of the lower energy cosmic rays are believed to be due to the explosion of stars into the normal interstellar medium (Lagage & Cesarsky 1983) and some of the higher energy cosmic rays to explosions of massive stars into their stellar wind (Völk & Biermann, 1988). For the energy range beyond a few $10^{15}$ eV there is general dispute about the origin of the particles.

Various models exist:
- 1. A postulated galactic wind (Jokipii & Morfill 1987) may accelerate particles at a galactic wind termination shock; this is argued to contribute particles over the entire range of particle energies, from low energies to the end of the cosmic ray spectrum.
- 2. The multiple shocks in the environment of OB superbubbles and young supernova remnants (Bykov & Toptygin 1990, 1992, Polcaro *et al.* 1991, 1993, Bykov & Fleishman 1992, Ip & Axford 1992) may also contribute.
- 3. The cosmic background of active galactic nuclei may contribute through the production of energetic neutrons which convert back to

- protons (Protheroe & Szabo 1992). This is a model to account mainly for the sharpness of the spectral turnover near a few $10^{15}$ eV.

- 4. Many authors have tried time and again to associate the the high energy cosmic rays with pulsars (a relevant short discussion is given by Hillas 1984).

For all such models, a clear prediction of the spectrum and its chemical abundance distribution, as well as a detailed check with the airshower size distribution, both for slanted and oblique showers, is desirable.

Recently we have proposed that the origin of cosmic rays can be explained as arising from three sources:

- 1. Supernova explosions into the interstellar medium,

- 2. Supernova explosions into a predecessor stellar wind,

- 3. The hot spots of giant powerful radiogalaxies.

The theory makes specific and quantitative predictions as to the origin of the energies of the particles, to the spectrum, and to the chemical composition.

The bridge in particle energies between the well established notion (due to Baade & Zwicky 1934, Ginzburg 1953, and Shklovsky 1953) that normal supernovae and supernova remnants provide some of the cosmic ray particles, and the other old idea (due to Cocconi 1956) that powerful extragalactic sources provide the extremely high energies, is the concept of stellar wind supernovae. Explosions into stellar winds have several clear differences to explosions into the interstellar medium: 1) The explosion is into a density law which decreases outwards rather than into a medium of constant density. 2) As a consequence the shock velocity stays high rather than to decrease steadily. 3) The magnetic field in stellar winds - required to explain nonthermal radioemission in OB stars, Wolf Rayet stars and radio supernovae - can be much higher than in the interstellar medium. 4) Prior to the explosion the winds in massive stars are observationally known to be strongly enriched in heavy chemical elements. 5) The massive stars with strong stellar winds are known to give about 1 in 4 supernovae.

The theory has been developed, tested and described in Biermann (1993a), Biermann & Cassinelli (1993), Biermann & Strom (1993), Stanev *et al.* (1993), Rachen & Biermann (1993), Rachen *et al.* (1993), Nath & Biermann (1993), Nath & Biermann (1994a, 1994b), and in Biermann (1993b, 1993c, 1993d, 1994). We will briefly review the predictions and some of the main tests.

However, we will start by discussing the key assumptions inherent in the model.

## 2. Key assumptions

### 2.1. THE TRANSPORT OF ENERGETIC PARTICLES IN THE ACCELERATION REGION

We argue that the cosmic ray components of nuclei from Helium and higher as well as all energetic electrons above about 30 GeV arise from supernova explosions into stellar winds. Check are the nonthermal radio emission (Abbott *et al.* 1986, Bieging *et al.* 1989) observed from stars with powerful winds, and the gamma-ray line emission observed by COMPTEL in the Orion region (Nath & Biermann, 1994b; see below).

The normal asymptotic configuration for an embedded magnetic field in a stellar wind has been derived by Parker (1958), and gives a magnetic field, which is tangential, decreases with radius $r$ as $1/r$, and with colatitude $\theta$ as $\sin \theta$. Consider a spherical shock wave in such a wind. We will use the approximation throughout, that the shock wave is spherically symmetric, and that any asymmetry is introduced by the orientation and latitude dependence of the magnetic field only. This is a strong simplification, but is necessary to keep the issue clear which we discuss here, the physics of shock waves which are at least locally spherical, and which run through a stellar wind. Then a spherical shock wave, centered on the star in its symmetry, compresses the magnetic field, this being tangential, by the full compression factor, in an adiabatic gas of index 5/3 by a factor of 4 for a strong shock.

Particles can be injected into an acceleration process (see, *e.g.*, Jones & Ellison 1991), and give a large proportion of the overall pressure and energy density. Then we have the case of a cosmic ray modified shock wave, such as treated by Zank *et al.* (1990). In the unperturbed state we have here a configuration, where the magnetic field is perpendicular to the shock direction, *i.e.*, parallel to the shock surface. Zank *et al.* demonstrated that for this case the configuration is neutrally stable, with an increasing instability as soon as the shock becomes more oblique relative to the underlying magnetic field, to lead to maximum instability for the case of a parallel shock configuration. In the oblique shock configuration, as many as three instabilities may operate (see their Table 1). In the case, which we consider here, this means that the slightest perturbation of the shock surface relative to the underlying magnetic field is leading to an instability. Such an instability then increases the deformation of the surface, which in turn strengthens the instability. Ultimately, the shock surface is maximally deformed, and changes its shape continuously, since the configuration is stuck in an unstable mode. Thus, strong turbulence is expected in the shock region, including downstream (Ratkiewicz *et al.* 1994).

Such a picture then leads to a concept, where the shock surface is jumping around an average location (a sphere in our case), where the upstream length scale of this jumping is a length corresponding to the same column densi-ty as the average downstream region, which corresponds to all the matter snowplowed in the expansion of the spherical shock. It is important to note that the length scales associated with the instability are hydrodynamic and therefore we are justified to use hydrodynamic scales below. In this concept the acceleration is a combination of a) the drifts the particles experience in the upstream or downstream regions, since the averaged magnetic fields are, of course, still perpendicular to the shock direction, and b) the energy gain from the Lorentz transformation each time a particle goes through the shock. At the same time, particles also lose energy from adiabatic expansion, since in the expansion of a spherical shock the local length scales always increase.

We have to caution, that we use here results from cosmic ray transport theory in an environment for which this theory was not made, *i.e.* where the magnetized ionized plasma is dominantly turbulent.

How can we test such a concept with observations? The ***radio observations*** of young supernova remnants and the nova GK Per can be a guide here. Radio polarization observations of supernova remnants clearly indicate what the typical local structure of these shocked plasmas is. Obviously, the normal expansion of a supernova remnant is not into a stellar wind, but into the interstellar medium. However, the typical magnetic field is highly oblique on average in a random field, and so the essential issue remains the same as in the case of a shock in a wind. The observational evidence (Milne 1971, Downs & Thompson 1972, Reynolds & Gilmore 1986, Milne 1987, Dickel *et al.* 1988) has been summarized by Dickel *et al.* (1991) in the statement that all shell type supernova remnants less than 1000 years old show dominant radial structure in their magnetic fields near their boundaries. There are several possible ways to explain this; we concentrate here on the idea, that this polarization pattern is due to rapid convective motion which induces locally strong shear.

These examples are for supernova explosions into the interstellar medium. There is also an observation demonstrating the same effect for an explosion into a wind: Seaquist *et al.* (1989) find for the spatially resolved shell of the nova GK Per a radially oriented magnetic field in the shell, while the overall dependence of the magnetic field on radial distance $r$ is deduced to be $1/r$ just as expected for the tightly wound up magnetic field in a wind. The interpretation given is that the shell is the higher density material behind a shock wave caused by the nova explosion in 1902 and now travelling through a wind. Seaquist *et al.* (1989) note the similarity to young supernova remnants.

The important conclusion for us here is that there appear to be strong radial differential motions in perpendicular shocks which provide the possibility that particles get ***convected*** parallel to the shock normal direction. The instability described earlier (Zank *et al.* 1990, Ratkiewicz *et al.* 1994) may be the physical reason for this rapid convective motion, we would like to suggest. We assume this to be a diffusive process, and note that others have also pointed out that this may be a key to shock acceleration (*e.g.* Falle 1990).

One supernova remnant rather shows a pattern in the motion of filaments: It appears that the radio knots in Cas A (Tuffs 1986) move erratically with a speed of the order of the shock speed itself in the expanding frame of the shock. This is actually important, it turns out, for our theory, as we use this erratic motion to limit the drift effect along the shock sphere.

This leads to the basic thesis underlying all the arguments made here: We propose that *a principle of the smallest dominant scale*, either in real space or in velocity space (Biermann 1993a; see Prandtl 1925 for a classical argument on such scaling), allows us to determine the relevant transport coefficients which describe the overall transport of particles in the shocked region, both parallel and transverse to the shock direction to derive drift energy gains and adiabatic losses.

## 2.2. THE TRANSPORT OF ENERGETIC PARTICLES IN THE GALAXY

There is a major problem with our approach (Biermann, 1993a): we take for the transport coefficient for cosmic ray particles out of the galactic disk a $E^{1/3}$ dependence, as a Kolmogorov spectrum of interstellar turbulence would imply, while a large body of evidence from the secondary to primary ratio in cosmic rays suggests that this dependence is closer to $E^{2/3}$ (Garcia-Munoz *et al.* 1987, Engelmann *et al.* 1990). While it may be possible that such a conclusion will have to changed using better cross sections for nuclear interactions (Webber *et al.* 1993), we will take this older, well established result on faith and consider its implications.

There is evidence from a variety of sources that the turbulence spectrum in the interstellar medium is of a Kolmogorov character. First, if one assumes that the turbulence in the interstellar medium responsible for scattering cosmic rays has a single powerlaw over the entire range of energies observed, then we have the limit that even at the highest particle energies the transit time through the disk should be larger than the simple light crossing time, because we observe no anisotropy. This means that the leakage time is weaker than $E^{-0.4}$ in its energy dependence. Second, the Chicago data already by themselves suggest that over a more limited particle energy range the exponent is close to $-1/3$ (Swordy *et al.* 1990). Third, all evidence from the interstellar medium itself (Larson 1979, 1981, Jokipii 1988, Narayan 1988, Rickett 1990) suggests a turbulence consistent with a Kolmogorov character: Here, the most definitive measurements come from Very Long Baseline Interferometry (VLBI) at radiowavelengths. With this technique one can observe water masers, pulsars and quasars (Blandford *et al.* 1986, Gwinn *et al.* 1988a, 1988b, Spangler & Gwinn 1990, Britzen 1993) and always finds properties of the interstellar medium consistent with the Kolmogorov spec-trum. Fourth, and finally, detailed plasma simulations demonstrate that in the regime where the magnetic field is of a strength similar to the thermal pressure the natural turbulence scaling is indeed Kolmogorov (Matthaeus & Zhou 1989), and only in the limit of strongly dominating magnetic field do we get a different behaviour (the Kraichnan law: Kraichnan 1965). Summarizing, it appears that a large body of evidence points to a Kolmogorov law. It should be pointed out, that with such a model for the escape of cosmic rays the air shower data for cosmic rays through the knee region as well as the electron spectrum can be modelled (Stanev *et al.* 1993, Rachen *et al.* 1993, Biermann 1994); the high energy electron spectrum constrains the escape time energy dependence also to a behaviour weaker than about $E^{-0.4}$ (Biermann 1994, and Wiebel, Biermann & Meyer, in prep.).

How can such contradictory evidence be reconciled? Many lines of evidence suggest that the linewidth of the molecular emission lines in clouds are far too broad to be understood as subsonic or even sonic turbulence. Supersonic turbulence, however, is far too dissipative to be reasonable. As a solution Alfvénic turbulence has been suggested (Genzel & Stutzki 1989). In that case trapping can occur during the formation of large molecular clouds and the escaping energetic particles experience interaction with matter proportional to $E^{-1/3}$, up to a critical particle energy where diffusion through the cloud is no longer a proper approximation; this can be shown analytically as well as with Monte-Carlo simulations (Stanev, Seo & Biermann, in prep.). The particles have a time scale to escape from the Galaxy with an energy dependence of again $E^{-1/3}$, and so the secondary particles spectrum seen is steeper by $E^{-2/3}$ than the primary spectrum in a first simple approximation. This is true below a critical particle energy of order $Z$ 20 GeV, above which the secondary to primary ratio should approach a $E^{-1/3}$ law; this is seen in the particle spectra for Boron and Carbon above particle energies of about 50 GeV (Wiebel, Biermann & Meyer, in prep.). One result is that the primaries observed experience a much lower matter column than previously believed; the spectral fits shown by Swordy *et al.* (1993) indeed show systematic deviations and are more readily reconcilable with straight powerlaws of slope -8/3 than with the curvature implied by strong matter interaction. Interaction with the magnetic field waves can, of course, also lead to additional acceleration (Dogiel & Sharov 1990), which we ignore in this simple argument.

There is another standard argument on the diffusion of cosmic rays, and that is the possible anisotropy. We wish to emphasize that there is no convincing evidence for any anisotropy anywhere below $3 \, 10^{18}$ eV (see, *e.g.*, Watson 1992), and so the isotropy data do not support a change in character for the transport of cosmic rays below this energy. Also, the observations of the interstellar medium do not suggest that there is any critical length scale short of the thickness of the disk harboring the cosmic ray population, which could be associated with the change in slope near a few $10^{15}$ eV.

Thus we propose to reconcile the cosmic ray information with what we know about the interstellar medium otherwise. The theory for the transport of cosmic rays in an unsteady interstellar medium remains to be worked in proper detail; the remarks above can only serve as a sketch of the concept.

Therefore we will assume in all following arguments that the correction from the injection to the observed spectrum of energetic particles is a change by -1/3 in the power law index.

## 3. Predictions and Tests

### 3.1. AIR SHOWER AND OTHER DATA

We predict for protons a spectrum of $E^{-2.75\pm0.04}$ (Biermann & Strom 1993); the Akeno data fit gives $E^{-2.75}$ (Stanev, Biermann & Gaisser 1993). We predict for Helium and heavier elements $E^{-2.67-0.02\pm0.02}$ below the knee; the Akeno data also here give a spectrum very close to prediction, of $E^{-2.66}$. We predict for the nuclei beyond the knee $E^{-3.07-0.07\pm0.07}$ (Biermann 1993c). The Akeno data give $E^{-3.07}$. The world data set of all good high energy data also gives this spectrum, as well as the cutoff near the predicted particle energy (Rachen, Stanev & Biermann 1993). The Fly's Eye data (Gaisser *et al.* 1993, Bird *et al.* 1993) demonstrate that the chemical composition switches rapidly from a heavy composition to a light composition near 3 EeV, as predicted already in 1990 (Biermann 1993d), based on earlier arguments by Biermann & Strittmatter (1987).

We predict the particle energies at the bend, or knee, and the energies of the various cutoffs; the test with the Akeno data gives fitted values for these numbers near prediction. From the fit the numerical values are rather strongly constrained, to within 20%, since we fit both vertical and slanted showers simultaneously in their showersize distribution.

We have difficulty estimating the systematic error resulting from the quite general use of limiting arguments, such as strong shocks, maximal curvature in the elements of the fast convection, and ignoring possible enhancements of the magnetic field strength (possibly important for the drifts) and cosmic ray pressure (except for the stability argument). The predictions and fits, which appear to agree generally quite well, illustrate the possible refinements required within the context of the approximations made.

Further checks are possible with i) the data analysis of Seo *et al.* (1991), who give a spectrum of $E^{-2.74\pm0.02}$ for Hydrogen, and $E^{-2.68\pm0.03}$ for Helium; ii) The analysis of Freudenreich *et al.* (1990) who have argued for some time, that the chemical composition near the knee becomes heavily enriched; iii) the newest Fly's Eye data (Bird *et al.* 1994) which give a spectrum of $E^{-3.07\pm0.01}$ beyond the knee in the energy range 0.2 to 80 EeV for the mono-ocular data, and $E^{-3.18\pm0.02}$ in the energy range 0.2 to 40 EeV for stereo data, while the classical data from Haverah Park (Cunningham *et al.* 1980; also see Sun *et al.* 1993) give a spectrum of $E^{-3.09\pm0.02}$ below 10 EeV; and the iv) JACEE data, presented at Calgary (Asakimori *et al.* 1993a, b), which also give spectra for Hydrogen and Helium at GeV particle energies, of $E^{-2.77\pm0.06}$ and $E^{-2.67\pm0.08}$, respectively.

### 3.2. THE GAMMA-RAY LINE EMISSION FROM THE ORION REGION

The COMPTEL instrument onboard GRO has detected gamma-ray line emission in the energy range 4 to 6 MeV, a spectral region where the continuum emission is very low, all coming from the star forming region Orion. In order to explain this, interaction of MeV energy cosmic rays with the interstellar medium is required; in any such model extremely enriched abundances of Oxygen and Carbon are implied. We showed (Nath & Biermann 1994b) that such lines arise naturally in the winds of OB and Wolf Rayet stars, when a large scale shock hits the wind shell of dense molecular gas, using normal abundances. The difference to the models of Polcaro *et al.* (1991, 1993), and Bykov & Toptygin (1990, 1992), as well as Bykov & Fleishman (1992) is that they consider the shock between the powerful stellar wind and the interstellar medium, while we consider a shock which travels down the wind and then interacts with the stand-off shock region of high density. Using the shock parameters given by optical data and their interpretation, we can at once fit the nonthermal radioemission in its spectrum and strength, the low gamma-ray continuum emission from $\pi^o$-decay, and the high gamma-ray line luminosity from excited energetic nuclei of Carbon and Oxygen. Just a few stars are required to explain the data. We consider this as a low energy test of the idea that shocks going through stellar winds accelerate nuclei.

### 3.3. THE ELECTRON SPECTRUM

At particle energies of order GeV, the radioemission of external galaxies is the most reliable indicator of the electron spectrum; Golla (1989) finds, that all excellent data (*a sample of seven galaxies*) are compatible with a single spectral index of $2.76 \pm 0.12$. This is very nearly the same spectral index as found for Hydrogen in cosmic rays. Wiebel (1992) compiled all the data available in the literature, and finds above about 30 GeV a spectrum of $E^{-3.26\pm0.06}$, which is to be compared with an expected spectrum from wind shocks of $E^{-3.33-0.02\pm0.02}$, steeper by unity than the injected spectrum due to synchrotron and inverse Compton losses. The particle energy of the switch between the two source sites, and the maximal electron energy observed, as well as the positron fraction, remain to be discussed in detail. The agreement is an important and independent test of the assumption of a Kolmogorov spectrum for the interstellar turbulence.

### 3.4. THE EXTRAGALACTIC SOURCES

The powerful radiogalaxies give a spectrum of cosmic rays at earth, after traversing the microwave background, which produces a sharp transition from a medium to heavy nuclei dominated chemical composition - due to stellar wind explosions - to a Hydrogen and Helium dominated composition near 3 EeV, as observed now by Fly's Eye; this had been predicted by the model a couple of years ago already (at Shapiro's meeting in 1990: Biermann 1993b), and also here the data can be fitted quite satisfactorily in particle energy, spectrum and chemical composition (Rachen & Biermann 1993, Rachen *et al.* 1993). The propagation of very high energy cosmic rays through the universe naturally leads to a minimum hypothesis to account for the topology of the intergalactic magnetic field: Galactic winds from normal and starburst galaxies carry out low energy cosmic rays as well as magnetic fields. While the low energy cosmic rays can reionize the intergalactic medium (Nath & Biermann 1993), the magnetic fields become very weak upon expansion and produce a topology which is commensurate with the large scale structure of the galaxy distribution. This then leads naturally to a mean free path for scattering of order 100 Mpc, which ensures that very high energy cosmic rays can come to us very nearly on straight paths, which is required to minimize their losses against the microwave background. The sky map of events gives important insight in a comparison with a map of powerful radio galaxies (Rachen *et al.*, in prep.).

### 3.5. THE HIGHEST ENERGY EVENTS

The most recent three extremely high energy events, beyond $10^{20}$ eV, from Fly's Eye, Akeno and Irkutsk can be explained with heavy nuclei in the context of our theory, provided that the energy estimate for the observed events is too high by about a factor of two (Rachen *et al.*, in prep.); this is quite easily possible due to the systematic errors inherent in the energy determination. On the other hand, should the energy estimates for these events be as high as published or even higher - systematic errors could go the other way also - then somewhat more exotic theories have to be sought to explain them (see, *e.g.*, Bhattacharjee 1991). Currently, all data available in public are consistent with the notion that powerful radiogalaxies are the sources of the cosmic ray particles above about 3 EeV.

## 4. Summary

Starting from a concept of the ***smallest dominant scale*** to describe the transport of energetic particles in a turbulent plasma in a cosmic ray mediated shock, we predict spectra for the galactic sites of cosmic ray origin. The approach is based on using limiting scales and so gives segments of powerlaws and cutoffs. For the cosmic rays we propose three sites of origin: i)

Explosions of supernovae into the interstellar medium. This provides for all Hydrogen observed and energetic electrons up to about 30 GeV. ii) Explosions of supernovae into stellar winds. This gives Helium and heavier elements as well as electrons above about 30 GeV. This component is heavily enriched beyond the knee, which is a feature in the spectra of cosmic rays accelerated in stellar winds due to a reduced drift energy gain. iii) Near 3 EeV there is a rapid switch in chemical composition to Hydrogen and Helium and the dominant source becomes the cosmological population of powerful radiogalaxies.

The model has passed a number of tests against observations, allowing first quantitative checks to be made. Further work will concentrate on the transport of cosmic rays, the secondary to primary ratio, and the farinfrared-radio emission correlation of galaxies.

## 5. Acknowledgements

I wish to thank my collaborators Drs. J.P. Cassinelli, H. Falcke, T.K. Gaisser, J.R. Jokipii, B. Nath, J. Rachen, T. Stanev, and R.G. Strom as well as Drs. V.S. Berezinsky, P. Bhattacharjee, J.H. Bieging, A. Bykov, P. Charbonneau, J.R. Dickel, V. Dogiel, P. Evenson, C. Jarlskog, J. Kirk, R.N. Manchester, J.F. McKenzie, H. Meyer, M. Nagano, S.P. Owocki, V.S. Ptuskin, J. Raymond, E.R. Seaquist, E.-S. Seo, F.D. Seward, V.V. Usov, H.J. Völk, G. Webb, and G. Zank for many intense discussions. High Energy Physics with the author has been supported by the Deutsche Forschungsgemeinschaft (DFG Bi 191/6, 7, 9), the Bundesministerium für Forschung und Technologie (DARA FKZ 50 OR 9202) and a NATO travel grant (CRG 9100072).